\documentclass[preprint]{revtex4-1}

\usepackage{graphicx}
\usepackage{dcolumn}
\usepackage{amsmath}
\usepackage{epstopdf}

\DeclareMathOperator{\Ai}{Ai}
\DeclareMathOperator{\Bi}{Bi}

\begin{document}


\title{Exact solutions for scalar field cosmology in f(R) gravity }



\author{S. D. Maharaj}\email[]{maharaj@ukzn.ac.za}
\author{R. Goswami}\email[]{vitasta9@gmail.com}
\affiliation{Astrophysics and Cosmology Research Unit, School of Mathematics, Statistics and Computer Science,
University of KwaZulu--Natal, Private Bag X54001, Durban 4000, South Africa.}
\author{S. V. Chervon}\email[]{chervon.sergey@gmail.com}
\affiliation{Astrophysics and Cosmology Research Unit, School of Mathematics, Statistics and Computer Science,
University of KwaZulu--Natal, Private Bag X54001, Durban 4000, South Africa.\\}
\affiliation{Laboratory of Gravitation, Cosmology, Astrophysics, Ilya Ulyanov State Pedagogical University, 100-years V.I. Lenin's Birthday square, Ulyanovsk 432700, Russia.}
\author{A. V. Nikolaev}\email[]{ilc@xhns.org}
\affiliation{Laboratory of Gravitation, Cosmology, Astrophysics, Ilya Ulyanov State Pedagogical University, 100-years V.I. Lenin's Birthday square, Ulyanovsk 432700, Russia.}

\date{\today}

\begin{abstract}
	We look for exact solutions in scalar field cosmology. To achieve this we use $f(R)$ modified gravity with a scalar field and do not specify the the form of the $f(R)$ function. In particular, we  study Friedmann universe assuming that acceleration of the scalar curvature is negligible. We first present solutions for special cases and then the general solution. Using initial conditions which represent the universe at the present epoch, we evaluated the constants of integration. This allows for the comparison of the scale factor in the new solutions with that of the $\Lambda CDM$ solution, thereby affecting the age of the universe in $f(R)$ gravity.
\end{abstract}


\maketitle


\section{Introduction}

When the evolution of the universe is analyzed we are faced with many types of scalar fields. During the inflationary stage in the early universe evolution we have the inflationary field -- the inflaton. When the current acceleration of the universe  started we consider as the reason a light scalar field termed quintessence. Dark energy, in the wide sense as the ingredient  responsible for the present universe acceleration, can be described as cosmological constant -- the  $\Lambda$CDM model, or alternatively as the $\phi$CDM or the $\sigma$CDM models which contain scalar fields in the form of canonical or phantom ones. The success of scalar field cosmology in the Friedmann universe based on general relativity gives a reason to investigate the influence of a self-interacting scalar field on universe evolution in cosmology based on $f(R)$ theory of gravity.

One of the most popular alternatives to standard general relativity is based on modifications of the standard Einstein-Hilbert action, by introducing terms which are nonlinear in the Ricci scalar $R$. Hence instead of the Ricci scalar $R$, the gravitational Lagrangian now becomes a well defined and at least $C^2$ function $f(R)$. We can easily see that general relativity is a special case of these class of theories with $f(R)=R$. It has been rigorously shown by many authors \cite{s1,u1,m1,zhuche00} that these modifications in the theory of gravity naturally admit a phase of late time accelerated expansion, as well as an early universe inflationary phase. In this way dark energy can be viewed as a geometrical quantity, rather than vacuum energy or additional scalar fields which are added by hand to the energy momentum tensor, to explain the recent cosmological observations. One of the important features of these $f(R)$ theories of gravity is that they can be always made to avoid instabilities that are inherent in higher order theories by judiciously choosing the parameters of the theory (or the function $f(R)$).

There have been several important studies on scalar field cosmology in general relativity and $f(R)$ gravity. The existence and stability of homogeneous and isotropic solutions in general relativity were established by Barrow and Ottewill \cite{barrow1} by considering perturbations into $f(R)$ theories. The cosmological and weak field properties of $f(R)$ theories were studied by Clifton and Barrow \cite{barrow2} with a Lagrangian $R^{1+ \delta}$; general
relativity is regained when $\delta \rightarrow 0$. The conformal structure of gravity in higher dimensions and inflation were examined in \cite{barrow3} which established a conformal equivalence to general relativity with a scalar field matter source. Astrophysical implications of inflating universes with quadratic curvature action were analyzed in \cite{barrow4} and inflationary models where the inflation was driven by existence of scalar fields were generated in \cite{barrow5}. Chakraborty and Sengupta \cite{chak} solved the field equations in higher order gravity by using the mathematical equivalence of these theories with scalar-tensor representations. The Noether symmetry approach \cite{rosh1,rosh2} and the Hojman conservation theorem \cite{rosh3} have been useful methods of finding solutions in $f(R)$ gravity with scalar fields. An exact cosmological solution in a covariant scalar-tensor-vector gravity was generated by Roshan \cite{rosh4}. Reviews of $f(R)$ theories and the methods of solving modified field equations with scalar field matter are contained in \cite{odin1,odin2,odin3}.

In this paper we find a number of theory independent exact solutions for the scalar field universe in higher order gravity, with the potential for the scalar field depending on the form of the function $f(R)$ chosen. We show that these solutions exactly mimic the standard $\Lambda$CDM cosmology in the near past (till at least $z=2$) and near future. This clearly indicates the degeneracy of admissible cosmological models that explains the recent cosmological observations in these higher order theories, due to the presence of extra scalar degrees of freedom, which is absent in general relativity.

\section{The model}
We start from the action
\begin{equation}
  S=\frac{1}{2\kappa^2}\int \sqrt{-g}d^4x \left\{ f(R) - \phi_{,\mu}\phi_{,\nu}g^{\mu\nu}-2V(\phi) \right\},
  \label{fre:act}
\end{equation}
where $\phi$ is the self-interacting scalar field with the stress energy tensor (SET)
\begin{equation}
  T_{\mu\nu} = \phi_{,\mu}\phi_{,\nu} + g_{\mu\nu}\left(\frac{1}{2}\phi^{,\alpha}\phi_{,\alpha} + V(\phi)\right),
  \label{fre:SET}
\end{equation}
and  $V(\phi)$ is the potential.

Our consideration will be in the spatially flat Friedmann universe (FLRW) with the metric
\begin{equation}
  ds^2=-dt^2+a^2(t)\left( dr^2 + r^2\left( d\theta^2 + \sin^2\theta  d\phi^2\right) \right)\;.
  \label{fre:frw}
\end{equation}
Einstein's equations in $f(R)$ gravity with SET~\eqref{fre:SET} for the metric~\eqref{fre:frw} are
the  Friedmann and Raychaudhury equations \cite{s1}
\begin{eqnarray}\label{fre:ens2}
  H^2=\frac{1}{3f'}(K+V) +\frac{R}{6}-\frac{f}{6f'}-H\frac{\dot{f}'}{f'},\\
  \dot{H} + H^2= -\frac{1}{3f'}(2K-V) -\frac{f}{6f'}+\frac{R}{6}-\frac{H\dot{f}'}{2f'}-\frac{\ddot{f}'}{2f'},
  \label{fre:ens1}
\end{eqnarray}
where $ K=\frac{1}{2} \dot{\phi}^2$ is the kinetic energy of the scalar field $ \phi$. A dot denotes differentiation with respect to time and the prime denotes differentiation with respect to the scalar curvature.
Einstein gravitational constant $\kappa^2$ is set equal to unity:  $\kappa^2=1$.
The scalar curvature $R$ for FRW metric (\ref{fre:frw}) is given as
$R=6(\dot{H}+2H^2)$.
Varying the action (\ref{fre:act}) with $\phi$ leads to the scalar field equation:
\begin{equation}
  \ddot\phi+3H\dot\phi+V'=0.
  \label{fre:fld}
\end{equation}
It is possible to express the kinetic energy $K$ and the potential $V$ from (\ref{fre:ens2}) and  (\ref{fre:ens1}):
\begin{eqnarray}
  \label{fre:kin}
  K &=& -f'\dot{H}+\frac{1}{2} H \dot{f}'-\frac{1}{2}\ddot{f}',\\
  V &=& -f' (2\dot{H}+3H^2)+\frac{5}{2}H\dot{f}'+\frac{1}{2} \ddot{f}'+\frac{1}{2}f.
  \label{fre:pot}
\end{eqnarray}
Substituting \eqref{fre:kin} and \eqref{fre:pot} in the field equation \eqref{fre:fld}, and using its representation in terms of the super potential $W=K+V$ \cite{zhuche00}
\begin{eqnarray}
  \label{nn:1}
  W \equiv K+V = -3f'(\dot{H}+H^2) + 3H\dot{f}' + \frac{1}{2}f,\\
  3H\dot\phi^2+\dot{W}=0,
  \label{nn:2}
\end{eqnarray}
leads to the identity
\begin{equation}
  \dot{f}-6f'(\ddot{H}+4H\dot{H})=f'[\dot{R} - 6(\ddot{H}+4H\dot{H})] =f'[6\ddot{H}+24H\dot{H})-6(\ddot{H}+4H\dot{H})] =0.
  \label{nn:3}
\end{equation}
Thus the field equation \eqref{fre:fld} is automatically satisfied when Einstein equations \eqref{fre:ens2} and  \eqref{fre:ens1} are solved. The situation is the same as in standard scalar field FRW cosmology~\cite{u1}.
Let us use the transformations below
\begin{eqnarray}
  \label{nn:4}
  \dot{f}=f'\dot{R},~~  \dot{f}' = f''\dot{R} =f''\left(6\ddot{H}+24H\dot{H}\right),\\
  \ddot{f'} = f'''\dot{R}^2+f''\ddot{R} = f'''\left(6\ddot{H}+24H\dot{H}\right)^2 + f''\left(6\dddot{H}+24\dot{H}^2+24H\ddot{H}\right).
  \label{nn:5}
\end{eqnarray}
By eliminating $\dot{f}$ with the expressions \eqref{nn:4} and \eqref{nn:5}, $K$ and $V$ can be  transformed to
\begin{equation}
  K=-f'\dot{H}+3f''(4H^2\dot{H}-3H\ddot{H}-\dddot{H} -4\dot{H}^2)-18f'''(\ddot{H}+4H\dot{H})^2,
  \label{fre:avn_k}
\end{equation}
\begin{equation}
  V=\frac{1}{2} f -f'(2\dot{H}+3H^2)+f''(3\dddot{H}+27H\ddot{H}+
  60H^2\dot{H}+12\dot{H}^2)+18f'''(\ddot{H}+4H\dot{H})^2.
  \label{fre:avn_v}
\end{equation}
Let us analyze the acceleration
\begin{equation}
  \frac{\ddot{a}}{a}= H^2+\dot{H}.
  \label{nn:6}
\end{equation}
Using \eqref{fre:ens1} we obtain
\begin{equation}
  \frac{\ddot{a}}{a}=-\frac{2K-V}{3f'}-18 \frac{f'''}{f'}(\ddot{H} +4H\dot{H})^2-\frac{f''}{f'}\left( 15H \ddot{H} +12H^2 \dot{H}+3\dddot{H} +12\dot{H}^2 \right)-\frac{f}{6f'}+\dot{H}+2H^2.
  \label{fre:avn_aa}
\end{equation}
Then by substitution of \eqref{fre:avn_k} and \eqref{fre:avn_v} in \eqref{fre:avn_aa}, using the expression
\begin{equation}
  2K -V = -\frac{1}{2}f + 3f'H^2 - f''(9\dddot{H} + 45H\ddot{H} + 36\dot{H}^2 + 36H^2\dot{H}) - 54f'''(\ddot{H}+4H\dot{H})^2,
  \label{nn:7}
\end{equation}
we recover the relation \eqref{nn:6} once again.
This result means that system of equations are consistent and there is no non-trivial restriction on the function $f(R)$.

\section{Exact solutions}
We will search for exact solutions starting from tight restrictions on the spacetime and subsequently weakening them step by step.
As we know
\begin{equation}
  R = 6(\dot{H}+2H^2),
  \label{nn:8}
\end{equation}
and hence
\begin{equation}
  \dot{R} = 6(\ddot{H}+4H\dot{H}),
  \label{nn:9}
\end{equation}
and
\begin{equation}
  \ddot{R} = 6(\dddot{H}+4\dot{H}^2+4H\ddot{H}).
  \label{nn:10}
\end{equation}
We consider the spacetime with $\ddot{R}=0$.
This means that we are searching for the solution of the equation
\begin{equation}
  4H\ddot{H}+\dddot{H} + 4\dot{H}^2=0.
  \label{fre:meq}
\end{equation}
The evident solution of this equation is
\begin{equation}
  \dot{H}=0,
  \label{nn:12}
\end{equation}
which leads to de Sitter spacetime. Instead of solving \eqref{fre:meq}, we can solve \eqref{nn:9}
\begin{equation}
	6(\ddot{H}+4H\dot{H}) = const.
	\label{fre:meq2}
\end{equation}
By standard substitution $ \dot{H}=y(H), x=H, \ddot{H}=y\frac{dy}{dx}$ in \eqref{fre:meq2}
we can obtain the equation
\begin{equation}
  y'-\frac{A}{y}+4x=0.
  \label{nn:15}
\end{equation}
Before the considering of general solution for \eqref{nn:15} we discuss solutions for special values of constants.

\subsection{Case $\dot{R}=0$}
Let us analyze the case when $\dot{R}=A=0$.
Then from \eqref{nn:9} we have
\begin{equation}
  \ddot{H}+4\dot{H}H=0.
  \label{nn:16}
\end{equation}
In terms of $y-x$ we have
\begin{equation}
  yy'+4yx=0.
  \label{nn:17}
\end{equation}
The solution $y=0$  corresponds to inflation $H=const$. As we mentioned above, this correspondence is valid for the case $\ddot{R}=0$ also.
The second possibility is the equation
\begin{equation}
  y'+4x=0.
  \label{nn:18}
\end{equation}
The equation may be represented with \eqref{nn:8} in the form
\begin{equation}
  \frac{1}{6}\frac{d}{dt}R=\frac{d}{dt}(\dot{H}+2H^2)=0.
  \label{nn:19}
\end{equation}
Then the solution is
\begin{equation}
  \frac{R}{6}=\dot{H}+2H^2=B,~ B=const.
  \label{fre:solB}
\end{equation}
We mention that the solution \eqref{fre:solB} means that $ R=6B=constant $ which also represents the de Sitter universe. Also with the help of \eqref{fre:solB} it is easy to simplify
the kinetic energy \eqref{fre:avn_k} and the potential \eqref{fre:avn_v} to the following forms:
\begin{eqnarray}
  K=-f'\dot{H} = -f'(B-2H^2),\\
  V=\frac{1}{2} f-f'(2\dot{H}+3H^2)=\frac{1}{2} f-f'(2B-H^2).
  \label{fre:potential}
\end{eqnarray}
Let us consider various possibilities with the constant $B$.
\subsubsection{$B=0$}
For the case $B=0$ we have the solution
\begin{equation}
  H=\frac{1}{2\tau},~~ a=a_* \tau^{1/2},~~\tau =(t-t_*)>0,
  \label{fre:a_sol1}
\end{equation}
where $t_*$ is an integration constant. Note that a similar solution for an FLRW radiation dominated universe was obtained in \cite{barrow1} when establishing the stability of general relativistic cosmological models.
As one can see this solution corresponds to radiation dominated universe.
We emphasise that the obtained regime is valid for any type of functional choice for $f$. The form of $f$ will do influence to kinetic and potential energy.
The kinetic energy will be determined by the relation
\begin{equation}
  K_I=\frac{f'}{2\tau^2}.
  \label{nn:20}
\end{equation}
Using the definition
\begin{equation}
  K=\frac{1}{2} \dot{\phi}^2,
  \label{nn:21}
\end{equation}
we can write
\begin{equation}
  \pm\phi = \int \sqrt{f'}(t-t_*)^{-1}dt.
  \label{nn:22}
\end{equation}
Further integration may be performed if we know the form of $f$, which is a function of $R$, i.e. a function on $t$.
Similar situation with the potential, which may be expressed as the function on time
\begin{equation}
  V=\frac{1}{2} f -\frac{f'}{4\tau^2}.
  \label{nn:23}
\end{equation}
For standard Friedmann cosmology we obtain an exponential potential as for any power law expansion of the universe:
\begin{equation}
  V(\phi)=H^2=\frac{1}{4} \exp(\mp 2(\phi-\phi_*).
  \label{nn:24}
\end{equation}
The scalar field has a logarithmic dependence on time
\begin{equation}
  \phi=\pm \ln \tau + \phi_*,~~ \phi_*=const.
  \label{nn:25}
\end{equation}
From \eqref{fre:solB} it is clear, that $ \frac{\ddot{a}}{a}=\dot{H}+H^2=-H^2<0$.
Thus the solution \eqref{fre:a_sol1} corresponds to a decelerating universe.

\subsubsection{$B<0$}

We now consider the case for $B<0$ .
Let $B=-\mu^2,~~ \mu=|\mu|$.
Integration gives us the solution
\begin{eqnarray}
  H=\frac{\mu}{\sqrt{2}}\tan(\sqrt{2}\mu(t_*-t)),\\
  a=a_*|\cos (\sqrt{2}\mu(t_*-t))|^{1/2},
  \label{fre:a_sol2}
\end{eqnarray}
with the restriction on time
\begin{equation}
  |t_*-t|<\frac{\pi}{2}\frac{1}{\sqrt{2}\mu}.
  \label{nn:27}
\end{equation}
Let us consider the acceleration in this case. From eq. \eqref{fre:solB} we can easily find
\begin{equation}
  \frac{\ddot{a}}{a}=\dot{H}+H^2=-\mu^2-H^2.
  \label{nn:28}
\end{equation}
This solution is not interesting because it has no inflationary stage.

\subsubsection{$B>0$}

Let us analyze the case $B>0$
Let $B=\nu^2, ~~\nu=|\nu|$.
Here two subcases arise, namely $H^2>\frac{\nu^2}{2} $ and $H^2<\frac{\nu^2}{2} $.

\textbf{In the first case}
$H^2>\frac{\nu^2}{2} $
the eq. \eqref{fre:solB} gives
\begin{equation}
  \frac{\ddot{a}}{a}=\dot{H}+H^2=-H^2+\nu^2,
  \label{nn:29}
\end{equation}
The restriction
$H^2>\frac{\nu^2}{2} $
leads to the solution
\begin{eqnarray}
  H=\frac{\nu}{\sqrt{2}}\coth(\sqrt{2}\nu(t+t_{pl})),\\
  a=a_*(\sinh (\sqrt{2}\nu(t+t_{pl})))^{1/2}.
  \label{fre:a_sol4}
\end{eqnarray}
We observe that potentials of the above form also arises in the analysis of Barrow \cite{barrow5}, when considering inflationary cosmology driven by scalar fields in general relativity. We can interpret our result as a generalization since this has been derived in higher order gravity.
The behaviour of \eqref{fre:a_sol4} is shown in Figure \ref{figure4_0}.%
\begin{figure}[!htb]
  \centering
  \input{figure3_0.tex}
  \caption{Plot of \eqref{fre:a_sol4}}
  \label{figure4_0}
\end{figure}
Let us analyze the solution \eqref{fre:a_sol4}.
Let the initial conditions be $a(0)=a_{pl}$, i.e. the universe has the Planck size at $t=0$. Also we can put as usual $a(t_0)=a_0=1$ at the present time. Then
\begin{eqnarray}
  a_{pl} = a_*(\sinh(\sqrt{2}\nu(t_{pl})))^{1/2,}\\
  a_0 = a_*\sinh^{1/2}(\sqrt\nu t_0) = 1.
  \label{nn:32}
\end{eqnarray}
If we substitute $\dot{H}$
\begin{equation}
  \dot{H} = -\nu^2\sinh^{-2}(\sqrt{2}\nu(t+t_{pl})),
  \label{nn:33}
\end{equation}
in\eqref{fre:potential} we find
\begin{eqnarray}
  K = f'\nu^2\sinh^{-2}(\sqrt{2}\nu(t+t_{pl})),\\
  V = \frac{1}{2}f-f'(-2\nu^2\sinh^{-2}(\sqrt{2}\nu(t+t_{pl}))+3\frac{\nu^2}{2}\coth^2(\sqrt{2}\nu(t+t_{pl}))).
  \label{nn:35}
\end{eqnarray}
As we can see the acceleration $\frac{\ddot{a}}{a}=H^2+\dot{H}=-H^2+\nu^2 =0$ may change the sign for the solution \eqref{fre:a_sol4} at the time when $ \coth^2(\sqrt{2}\nu(t+t_{pl}))= 2$ and the deceleration will be changed to the acceleration of the universe. So this solution may correspond to the late time accelaration.

\textbf{In the second case}
$H^2<\frac{\nu^2}{2} $
and the universe is an expanding one. Moreover it is clear that inequality may be extended to $H^2<\frac{\nu^2}{2}<\nu^2.$ The last condition gives the existence of
acceleration in the expanding universe. Thus the solution for this
case may describe an inflationary phase without exit.
The solution is
\begin{eqnarray}
  H=\frac{\nu}{\sqrt{2}}\tanh(\sqrt{2}\nu(t+t_{pl}))\,\
  a=a_*(\cosh (\sqrt{2}\nu(t+t_{pl})))^{1/2}.
  \label{fre:a_sol3}
\end{eqnarray}
Let us analyze the solution \eqref{fre:a_sol3}. At first step let us use the initial conditions $a(0)=a_{pl}$, i.e. the universe has the Planck size at $t=0$. Then 
\begin{equation}
  a_{pl} = a_*\cosh^{1/2}(\sqrt{2}\nu(t_{pl})),
  ~~a(t_{pl})=\left(\frac{\cosh (2\sqrt{2}\nu t_{pl})}{\cosh (\sqrt{2}\nu t_{pl})}\right)^{1/2},
  \label{nn:37}
\end{equation}
and we may connect the parameter $a_*$ with $a_{pl}$ and consider $\nu $ as a free parameter. Also we can put as usual $a(t_0)=a_0=1$.	
The behaviour of \eqref{fre:a_sol3} is shown on Figure \ref{figure5}.
\begin{figure}[!htb]
  \centering
  \input{figure4.tex}
  \caption{Plot of \eqref{fre:a_sol3}}
  \label{figure5}
\end{figure}
Using the derivative $\dot{H}$
\begin{equation}
  \dot{H} = \nu^2\cosh^{-2}(\sqrt{2}\nu(t+t_{pl})),
  \label{nn:38}
\end{equation}
from \eqref{fre:potential} we can find
\begin{eqnarray}
  K = -f'\nu^2\cosh^{-2}(\sqrt{2}\nu(t+t_{pl})),\\
  V = \frac{1}{2}f-f'\nu^2(2-\frac{1}{2} \tanh^2 (\sqrt{2}\nu(t+t_{pl}).
  \label{nn:40}
\end{eqnarray}
Thus we have nonsingular universe corresponding to the inflationary stage.
Let us note that in the case when $f'>0$ we have to consider
a phantom scalar field, because $K=\frac{1}{2} \dot{\phi}^2<0$ in \eqref{nn:38}.
Let as analyze the standard Friedmann universe with $f(R)=R$, which is general relativity.
Then we can find the solution for the phantom scalar field
\begin{equation}
  \phi=\pm 2 \arctan (\exp (\sqrt{2}\nu(t+t_{pl}))).
  \label{nn:41}
\end{equation}
After simplification, the potential can be reduced to the form
\begin{equation}
  V(\phi)=\nu^2 \left( 1+\frac{1}{2 \cos^2 \phi}\right).
  \label{nn:42}
\end{equation}
It should be mentioned that for all solutions the potential and kinetic energy may depending on $f$ and $f'$ only, not from higher orders of derivative.

\subsection{$A \ne 0$}

Let us return to the general equation \eqref{fre:meq}:
\begin{equation}
  \frac{1}{6}\dot{R}=\frac{d}{dt}\left( \dot{H}+2H^2\right)=A.
  \label{nn:44}
\end{equation}
Performing the integration we obtain
\begin{equation}
  \dot{H}+2H^2 =At+B.
  \label{nn:45}
\end{equation}
By transformation $U=2H $ we can avoid the multiplier 2:
\begin{equation}
  \dot{U}+U^2=2At+2B.
  \label{nn:46}
\end{equation}
Let us introduce new variable
\begin{equation}
  \xi =t+\frac{2B}{2A}.
  \label{nn:47}
\end{equation}
Then equation transforms to
\begin{equation}
  U_\xi +U^2=2A\xi.
  \label{nn:48}
\end{equation}
Then the equation transforms to the Airy equation
\begin{equation}
  \Theta_{\xi\xi}=2A\xi \Theta(\xi),
  \label{nn:49}
\end{equation}
where $\Theta = a^2$.
The solution for this equation by substitution $U=\frac{\Theta_\xi}{\Theta}$ is
\begin{equation}
  a^2(\xi)=C_1 \Ai\left( -(-2A)^{1/3}\xi\right)+C_2 \Bi\left( -(-2A)^{1/3}\xi\right),
  \label{fre:solAiBi}
\end{equation}
where $\Ai$ and $\Bi$ are the Airy functions of first and second kind.

\section{Finding constant of Integrations}

We will now try to evaluate the constants of integration with the aid of the different parameters of the universe evaluated at the present epoch.
\subsection{Set of initial conditions}
For the initial conditions we set the following:
\begin{itemize}
  \item $t_0 = 0$ for the present epoch.
  \item $a_0 = 1$ for scale factor at the present epoch, that is
    \begin{equation}
      a(0) = 1.
      \label{avn:ic1}
    \end{equation}
  \item Hubble factor for the present epoch defined by
    \begin{equation}
      H(0) = \left( \frac{\dot{a}}{a} \right)_{t=0} = H_0.
      \label{avn:ic2}
    \end{equation}
  \item Deceleration parameter at the present epoch is defined as
    \begin{equation}
      q(0) = \left( -\frac{\ddot{a}a}{\dot{a}^2} \right)_{t=0} = q_0.
      \label{avn:ic3}
    \end{equation}
\end{itemize}

\subsection{Standard $\Lambda CDM$ solution for the late time inflation}

We assume flat universe $k=0$ and absence of radiation $\Omega_R = 0$.
From the Friedmann equations
\begin{eqnarray}
  \frac{\ddot{a}}{a} = -\frac{4\pi G}{3}(\rho + 3p) ,\\
  \frac{\dot{a}^2}{a^2} = \frac{8\pi G \rho}{3},
  \label{avn:1}
\end{eqnarray}
and setting $p_\Lambda = - \rho_\Lambda$ and $p_M = 0$ we obtain 
\begin{equation}
  \frac{\ddot{a}}{a}=-\frac{1}{2}\frac{\dot{a}^2}{a^2} + \frac{3}{2}H_\Lambda^2,
  \label{anv:2}
\end{equation}
where
\begin{equation}
  H_\Lambda = \left(\frac{8\pi G\rho_\Lambda}{3}\right)^\frac{1}{2}.
  \label{avn:3}
\end{equation}
Solution for \eqref{avn:3} is
\begin{equation}
  2a^\frac{3}{2}H_\Lambda e^{-\frac{3}{2}H_\Lambda t} - e^{-3 H_\Lambda t}C_1 + C_2 = 0.
  \label{avn:4}
\end{equation}
 Let us calculate other parameters by using \eqref{avn:4}:
\begin{equation}
  H = \frac{H_\Lambda\left( e^{3H_\Lambda t}\left( C_1 - 2H_\Lambda \right) + C_1 \right)}{e^{3H_\Lambda t} \left( C_1 - 2H_\Lambda \right) - C_1},
  \label{avn:5}
\end{equation}
and
\begin{equation}
  q = -\frac{C_1^2}{\left( e^{-3H_\Lambda t}C_1 + C_1 - 2H_\Lambda \right)^2}e^{-6H_\Lambda t} - \frac{\left( -4 C_1^2 + 8 C_1 H_\Lambda \right)e^{-3H_\Lambda t} + C_1^2 - 4C_1H_\Lambda + 4H_\Lambda^2}{\left( e^{-3H_\Lambda t}C_1 + C_1 - 2H_\Lambda \right)^2}.
  \label{anv:6}
\end{equation}
Now we are ready to apply our set of initial conditions.
From \eqref{avn:ic1} we have
\begin{equation}
  C_2=-2H_\Lambda + C_1.
  \label{avn:7}
\end{equation}
From \eqref{avn:ic2} we have
\begin{equation}
  -C_1 + H_\Lambda = H_0.
  \label{avn:8}
\end{equation}
From \eqref{avn:ic3} we have
\begin{equation}
  \frac{C_1^2 - 2C_1 H_\Lambda - 2H_\Lambda^2}{2\left( C_1 - H_\Lambda \right)^2} = q_0.
  \label{avn:9}
\end{equation}
And from \eqref{avn:8} and \eqref{avn:9} we get
\begin{eqnarray}
  H_\Lambda & = & \frac{\sqrt{-6q_0 + 3}H_0}{3} \\
  C_1 & = & \sqrt{-\frac{2q_0}{3} + \frac{1}{3}}H_0 - H_0.
  \label{avn:10}
\end{eqnarray}
Thus as a result \eqref{avn:4} transforms to
\begin{equation}
	a(t) = \frac{1}{6} \left( -\frac{18\left(q_1 e^{q_1H_0 t} + 3e^{q_1H_0 t} + q_1 - 3 \right)^2 e^{-q_1H_0 t}}{2q_0 - 1} \right)^{1/3},
	\label{avn:11}
\end{equation}
where $q_1=\sqrt{-6q_0 + 3}$. It is also easy to see that this solution, using \eqref{avn:10} and definition of $H_\Lambda$ for known $\Omega_\Lambda$, $q_0$ can be found as 
\begin{equation}
  q_0 = \frac{H_0^2 - 3 H_\Lambda^2}{2H_0^2} = -\frac{3}{2}\Omega_\Lambda + \frac{1}{2}.
  \label{anv:12}
\end{equation}
For example for $\Omega_\Lambda = 0.72$ we have $q_0 = -0.58$.

\subsection{Solution \eqref{fre:a_sol4}}
We are starting from solution \eqref{fre:a_sol4}
\begin{equation}
  a = a_* \sinh^{1/2}\sqrt{2}\nu\left( t + t_* \right).
  \label{avn:13}
\end{equation}
Let us calculate other parameters by using \eqref{avn:4}
\begin{equation}
  H = \frac{\sqrt{2}\nu}{2}\coth{\sqrt{2}\nu\left( t + t_* \right)},
  \label{avn:14}
\end{equation}
and
\begin{equation}
  q = - \frac{\cosh^2\sqrt{2}\nu\left( t + t_* \right) - 2}{\cosh^2\sqrt{2}\nu\left( t+t_* \right)}.
  \label{anv:15}
\end{equation}
Now we are ready to apply our set of initial conditions.
From \eqref{avn:ic1} we have
\begin{equation}
  a_* = \frac{1}{\sqrt{\sinh{\sqrt{2}\nu t_*}}}.
  \label{avn:16}
\end{equation}
From \eqref{avn:ic2} we have
\begin{equation}
  \frac{\sqrt{2}\nu}{2}\coth{\sqrt{2}\nu\left( t_* \right)} = H_0.
  \label{avn:17}
\end{equation}
From \eqref{avn:ic3} we have
\begin{equation}
  - \frac{\cosh{\sqrt{2}\nu\left( t_* \right)}^2 - 2}{\cosh{\sqrt{2}\nu\left( t_* \right)}^2} = q_0.
  \label{avn:18}
\end{equation}
And from \eqref{avn:17} and \eqref{avn:18} we have
\begin{eqnarray}
  \nu & = &\frac{1}{4}\sqrt{2}H_0 Q, \\
  t_* & = &\frac{1}{H_0}\ln\left( -\frac{-2\sqrt{-2q_0+2}+q_0 -3}{1+q_0} \right)Q^{-1},
  \label{avn:19}
\end{eqnarray}
where
$$
Q = q_0 + \frac{-q_0^2 + 2q_0+2\sqrt{-2q_0 +2}q_0 +3 + 2\sqrt{-2q_0+2} }{1+q_0}-3.
$$Thus \eqref{avn:13} transforms to
\begin{equation}
  a(t) = \sinh^{-1/2}\left( \frac{1}{2}\ln{Q_1}\right) \sinh^{1/2}\left( \sqrt{-2q_0+2}H_0 t + \frac{1}{2}\ln{Q_1} \right),
  \label{avn:20}
\end{equation}
where
$$
Q_1 = \frac{-q_0+3+2\sqrt{-2q_0 + 2}}{1+q_0}.
$$

\subsection{Solution \eqref{fre:solAiBi}}

Let us rewrite equation \eqref{fre:solAiBi} in terms of the scale factor
\begin{equation}
  a = \sqrt{C_1\Ai\left(-(-2A)^{1/3}\left( t+ \frac{B}{A} \right)\right) + C_2\Bi\left(-(-2A)^{1/3}\left( t+ \frac{B}{A} \right)\right)}.
  \label{avn:21}
\end{equation}
Since the function $\Bi(t)$ diverges at the present epoch,  we can set up $C_2 = 0$ for physical reasons, and then
\begin{equation}
  a = \sqrt{C_1\Ai\left(-(-2A)^{1/3}\left( t+ \frac{B}{A} \right)\right)}.
  \label{avn:22}
\end{equation}
Let us calculate other parameters by using \eqref{avn:22}
\begin{equation}
  H = -\frac{(-2A)^{1/3}\Ai'\left( \xi \right)}{2\Ai\left( \xi \right)},
  \label{avn:23}
\end{equation}
and
\begin{equation}
  q = -2^{1/3}\frac{4\left( At+B \right)\Ai(\xi)^2 - 2^{2/3}(-A)^{2/3}\Ai'(\xi)^2}{2(-A)^{2/3}\Ai'(\xi)^2}.
  \label{avn:24}
\end{equation}
Now we are ready to apply our set of initial conditions.
From \eqref{avn:ic1} we have
\begin{equation}.
  C_1 = \Ai\left( \xi_0\right)^{-1}.
  \label{avn:25}
\end{equation}
From \eqref{avn:ic2} we have
\begin{equation}
  -2^{-2/3}(-A)^{1/3} \frac{\Ai'\left( \xi_0\right)}{\Ai\left( \xi_0\right)}= H_0.
  \label{avn:26}
\end{equation}
From \eqref{avn:ic3} we have
\begin{equation}
  -2^{-2/3}\frac{4B\Ai(\xi_0)^2 - 2^{2/3}(-A)^{2/3}\Ai'(\xi_0)^2}{(-A)^{2/3}\Ai'(\xi_0)^2}= q_0,
  \label{avn:27}
\end{equation}
where
\begin{equation}
  \xi_0 = 2^{1/3}B(-A)^{-2/3}.
  \label{avn:28}
\end{equation}
And from \eqref{avn:26} and \eqref{avn:27} we have
\begin{equation}
  B = -\left( q_0 - 1 \right)H_0^2,
  \label{avn:29}
\end{equation}
and constant $A$ could not be expressed in elementary functions but only as solution for the following equation
\begin{equation}
  A = -{\rm RootOf}\left( 2^{1/3}x \Ai'\left( -\frac{2^{1/3}\left( q_0 - 1 \right)H_0^2}{x^2} \right) + 2 \Ai\left(  -\frac{2^{1/3}\left( q_0 - 1 \right)H_0^2}{x^2} \right)H_0 \right)^3.
  \label{avn:30}
\end{equation}

 \begin{figure}[!htb]
  \centering
  \includegraphics[width=0.8\textwidth]{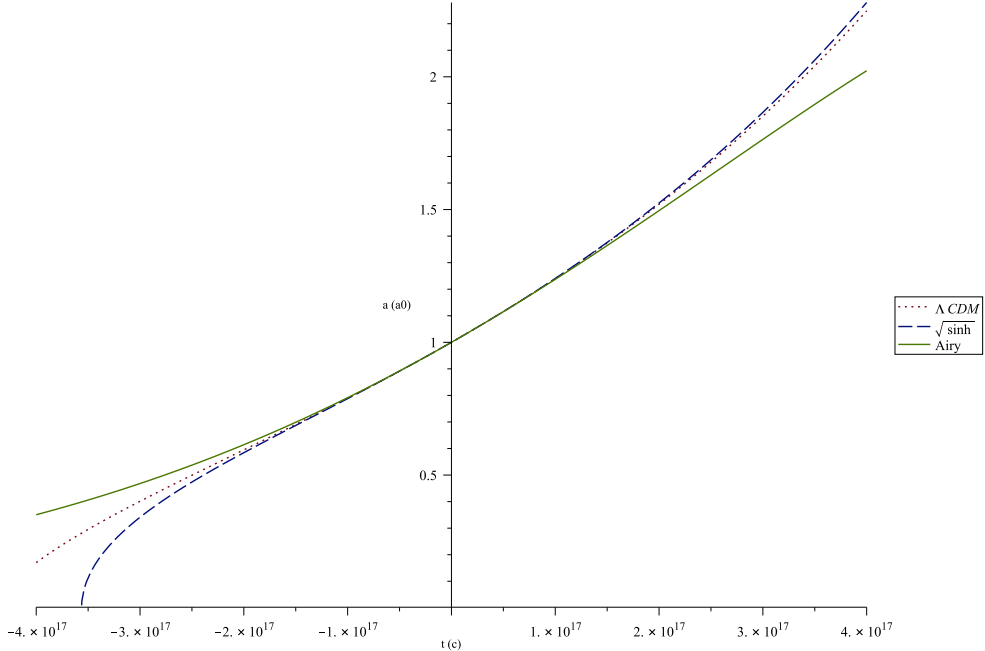}
  \caption{Plots for \eqref{avn:22}, \eqref{avn:20} and \eqref{avn:11} for $q_0 = -0.58$ and $H_0 = 69$}
\end{figure}

\section{Discussion}

From the above plots, which describes the scale factor evolution with respect to the cosmic time for the various solutions obtained in this paper, we have very interesting observations, which we state as follows:
\begin{enumerate}
\item At least till red shift $z=2$, the scalar field cosmologies in $f(R)$ gravity with constant or linearly varying Ricci scalar has remarkable similarities with the standard $\Lambda$CDM cosmology, in terms of the evolution of the scale factor. Hence by usual cosmological distance measurements (such as supernovae observations) till $z=2$, it is impossible to differentiate between these theories.
\item The deviation from the standard $\Lambda$CDM cosmology occurs in distant past (beyond $z=2$), in terms of the big bang epoch. In one of the solutions with constant Ricci scalar the big bang advances in time reducing the age of the universe, whereas for the solution with linearly varying Ricci scalar, the big bang is thrown further back, increasing the age of the universe.
\item The interesting points of these solutions are: they are true for all well behaved $f(R)$ theories and hence these are theory independent behaviours. The theory dependence creeps into the potential function of the scalar field, which will be different for different $f(R)$ theories.
\end{enumerate}

\section*{ACKNOWLEDGEMENTS}
SVC and AVN were partly supported by the State order of Ministry of education and science of RF number 2014/391 on the project 1670. RG is supported by National Research Foundation (NRF), South Africa. SDM acknowledges that this work is based on research supported by the South African Research Chair Initiative of the Department of
Science and Technology and the National Research Foundation.

\end{document}